\documentclass[twocolumn,english,aps,prb,floatfix,preprintnumbers,showpacs,amsfonts,amssymb]{revtex4}
\usepackage[T1]{fontenc}
\usepackage[latin1]{inputenc}
\usepackage{amsmath}
\usepackage{amssymb}

\makeatletter

\providecommand{\LyX}{L\kern-.1667em\lower.25em\hbox{Y}\kern-.125emX\@}

\usepackage{float}

\usepackage{graphicx}
\usepackage{graphicx}
\usepackage{graphicx}
\usepackage{amscd}
\usepackage{bm}

\usepackage{babel}
\makeatother
\begin{document}

\title{Dynamical localization of a particle coupled to a two-level systems
thermal reservoir}

\author{A. Villares Ferrer}

\affiliation{Instituto de F\'{\i}sica, Universidade Federal Fluminense, Av. General
Milton Tavares de Souza s/n, Gragoatá, Niter\'{o}i 24210-346, Rio
de Janeiro, Brazil.}

\author{C. Morais Smith}

\affiliation{Institute for Theoretical Physics, University of Utrecht, Leuvenlaan
4, 3584 CE, Utrecht, The Netherlands.}

\date{\today{}}

\begin{abstract}
Using the functional-integral method, we investigate the effect of
a two-level systems thermal reservoir on the single particle dynamics.
We find that at low temperatures, within the sub-ohmic regime, the
particle becomes {}``dynamically'' localized at long times due to an effective
potential generated by the particle-reservoir interaction. This behavior
is different from the one obtained for the usual bath of harmonic
oscillators and is fundamentally related with the non-Markovian
character of the dissipative process. 
\end{abstract}


\maketitle

\section{Introduction}

The influence of dissipation on quantum tunneling\cite{AOannal} and
on quantum coherence\cite{AJLrmodphy} has attracted much attention
during the last decades. In a pioneer work, Caldeira and Leggett\cite{AOannal}
suggested that a bosonic heat bath consisting of an infinite number
of harmonic oscillators constitutes an universal realization, which
can mimic a large variety of real environments.\cite{Weiss} However,
a real environment cannot \textit{always} be represented in this way.
Indeed, if the particle of interest couples to a non linear system,
the latter may, under certain circumstances, behave as a fermionic
heat bath. That is the case of a distribution of quartic plus quadratic
potentials, which within a very well known limit,\cite{quartic} effectively
acts as a collection of two-level systems (TLSs). In such a case the
environment truly behaves as a spin $1/2$ medium.

A spin-bath composed of an infinite number of TLSs has been mostly
considered when the system of interest is itself a TLS, \cite{PSrepprphy}
providing for instance, a realistic description of a nanomagnet coupled
to a set of surrounding spins\cite{MagTunn} and a useful model to
describe the loss of quantum coherence.\cite{Hangi,Makri} The representation
of dissipative environments by a spin-bath has been extended also
to situations in which the environment is not composed of real spins,
as e.g. in the description of disordered insulating solids\cite{twlinsu}
or for explaining the damping of acoustic phonons in a nanomechanical
resonator.\cite{ressonator} Moreover, it has been found that the
dissipative dynamics of a single particle linearly coupled to a TLSs
reservoir is non-Markovian\cite{twlAHCN} and that the transport properties
strongly differ from the usual oscillator thermal bath.\cite{twlyo}
Indeed, the optical conductivity of a set of non interacting particles
linearly coupled to a TLSs reservoir exhibits a remarkable non-Drude
behavior. In particular, in the sub-ohmic regime the system exhibits
a maximum in the incoherent electrical conductivity at finite frequency.\cite{twlyo}
This kind of behavior has been experimentally observed in La$_{2-x}$Sr$_{x}$CuO$_{4}$\cite{takenaka1}
and La$_{1-x}$SrMnO$_{3}$,\cite{takenaka2} in situations in which
inelastic scattering dominates the transport properties. The broad
finite energy peak observed experimentally suggests a \char`\"{}dynamical\char`\"{}
localization of the charged particles, similar to the non-Fermi-liquid
behavior found in the infrared conductivity of SrRuO$_{3}$.\cite{kostic}
In all the cases discussed above, the localization of the particle was 
attributed to inelastic scattering processes because they are strongly 
enhanced as the temperature is raised. This high-temperature localization 
is different from the Anderson localization, which tends to be destroyed 
by inelastic processes.

In this paper we present a simple model, which leads to {}``dynamical''
localization due to inelastic scattering, but at \textit{low temperatures}.
At this point, we want to call the attention of the reader to our
use of the term {}``dynamical'' localization. This term is conventionally
used in the literature\cite{dynamical} to designate quantum phenomena
taking place in time-periodic systems, whose corresponding classical
dynamics displays chaotic diffusion. The phenomenon described here
bears no similarities with the latter and for this reason we use the
term under quotation marks. The {}``dynamical'' localization
of the charge carriers studied here is generated by their coupling
to a TLSs thermal bath and is in close relation to a non-Markovian
process at the classical level. At first sight, a non-Markovian particle
dynamics, which implies that the particle never reaches equilibrium,
seems to be in contradiction with localization, which is characteristic
of insulators. This paper intends also to shed some light on this
point. In order to achieve our goal, we investigate the real time
effective dynamics of a single particle coupled to a TLSs thermal
reservoir in the sub-ohmic regime. The effective action, which describes
the particle interacting with the TLSs bath, is obtained using the
well known Feynman-Vernon formalism.\cite{FeyVer} 

This paper is divided as follows: In Sec. II
we present the model describing the particle of interest interacting
with the TLSs reservoir, as well as a brief sketch of the derivation 
of the effective particle dynamics. In Sec. III the ``dynamical'' localization
effect is explicitly derived and discussed in specific cases
within the sub-ohmic regime. Finally in Sec. IV we present our conclusions.

\section{The Model}

To begin with, we will describe the particle of interest coupled to
a generic TLSs thermal reservoir by the Hamiltonian \begin{equation}
H=\frac{{\hat{p}}^{2}}{2M}+u(x)+\sum _{k=1}^{N}\frac{{\hbar \omega _{k}}}{2}\sigma _{zk}-x\sum _{k=1}^{N}J_{k}\sigma _{xk},\label{eq:HTot}\end{equation}
 where the first two terms stand for a particle under the influence
of an arbitrary potential $u$, the third term accounts for the TLSs
reservoir, and the last one describes the interaction between the
particle and the thermal bath. $J_{k}$ denotes the coupling parameter
and $\sigma _{zk/xk}$ stand for Pauli matrices.

The first step is to calculate the reduced density operator of the
particle of interest, which may be obtained after tracing out the
reservoir degrees of freedom, \begin{equation}
\rho (x,y,t)={\textrm{Tr}}_{R}[\langle x|e^{-i\frac{{Ht}}{\hbar }}\rho (0)e^{i\frac{{Ht}}{\hbar }}|y\rangle ].\label{eq:res1}\end{equation}

The density operator of the total system at time $t=0$ will be assumed
to be decoupled, $\rho (0)=\rho _{S}(0)\rho _{R}(0)$. The reduced
density operator (\ref{eq:res1}) can then be written as \[
\rho (x,y,t)=\int dx'\int dy'\rho _{S}(x',y',0){\mathcal{J}}(x,y,t;x'y',0),\]
 where the super-propagator ${\mathcal{J}}$ has the form \begin{equation}
{\mathcal{J}}=\int _{x'}^{x}{\mathcal{D}}x(t')\int _{y'}^{y}{\mathcal{D}}y(t')e^{\frac{{i}}{\hbar }(S_{0}[x]-S_{0}[y])}{\mathcal{F}}[x,y].\label{eq:sup2}\end{equation}
 In the expression above, $S_{0}[x]$ corresponds to the action of
a free particle placed in the potential $u$, while ${\mathcal{F}}$
denotes the influence functional which describes the influence of
the reservoir on the particle dynamics. After integrating out the
reservoir degrees of freedom and introducing a set of coordinates
corresponding to the particle center of mass $q=(x+y)/2$ and relative
coordinate $\xi =x-y$, we obtain the super-propagator for the particle
of interest (see {[}\onlinecite{twlyo}{]} for details), \begin{equation}
\mathcal{J}[q,\xi ,t;q',\xi ',0]=\int _{\xi '}^{\xi }{\mathcal{D}}\xi \int _{q'}^{q}{\mathcal{D}}q\: e^{\frac{{i}}{\hbar }S_{\textrm{eff}}[q,\xi ]-\frac{{1}}{\hbar }\phi [\xi ]}.\label{eq:superpro1}\end{equation}
 The effective action is given by \begin{eqnarray}
S_{\textrm{eff}} & = & \int _{0}^{t}dt'\left[M\dot{q}(t')\dot{\xi }(t')-u(q,\xi )\right.\nonumber \\
 &  & \left.-\int _{0}^{t'}dt''\Lambda (t'-t'')q(t'')\xi (t')\right],\label{eq:effact}
\end{eqnarray}
 where \[
\Lambda =\int _{0}^{\infty }d\omega J(\omega ,T)\sin [\omega (t'-t'')],\]
 and the functional $\phi $ has the form \begin{equation}
\phi [\xi ]=\int _{0}^{t}dt'\int _{0}^{t'}dt''\Phi (t'-t'')\xi (t')\xi (t''),\label{eq:phifunctional}\end{equation}
 with\[
\Phi =\int _{0}^{\infty }d\omega J(\omega ,T)\, \cos [\omega (t'-t'')]\coth \left(\hbar \omega /2k_{B}T\right).\]
 It should be noticed that the kernels $\Lambda $ and $\Phi $ are
both defined in terms of the spectral density of the thermal reservoir\cite{twlyo}
$J(\omega ,T)$.

We now perform an integration by parts to render explicit the dependence
of the last term in Eq.\ (\ref{eq:effact}) on the velocity. We then
find \begin{eqnarray}
S_{\textrm{eff}} & = & \int _{0}^{t}dt'\left[M\dot{\xi }(t')\dot{q}(t')-\xi (t')\int _{0}^{t'}dt''{\tilde{\Gamma }}(t'-t'')\dot{q}(t'')\right.\nonumber \\
 & + & \left.{\tilde{\Gamma }}(0)\xi (t')q(t')-u(q,\xi )-q(0){\tilde{\Gamma }}(t')\xi (t')\right],\label{eq:seffexten}
\end{eqnarray}
 where \begin{equation}
{\tilde{\Gamma }}(t'-t'')=\int _{0}^{\infty }d\omega \frac{J(\omega ,T)}{\omega }\cos [\omega (t'-t'')].\label{eq:LambdaM}\end{equation}

Although in this form the effective action shows an explicit velocity
dependent term, characteristic of viscous forces, we have obtained
also two additional spurious contributions. The first term in the
second line of Eq.\ (\ref{eq:seffexten}) is nothing but a harmonic
potential, which can be canceled by an appropriate choice of the external
potential, \[
u(q,\xi )={\tilde{\Gamma }}(0)q\xi ={\tilde{\Gamma }}(0)(x²-y^{2})/2.\]
 This assumption allows us to focus on the dissipative effects of
the environment. An alternative procedure would be to start from a
momentum dependent particle-reservoir interaction. In this way the
effective action would immediately exhibit a velocity dependent term,
without any additional spurious contribution. To conclude the analysis
of Eq.\ (\ref{eq:seffexten}) we must point out that its last term
fluctuates very rapidly for times $t\gg 1/\omega $. Hence, in principle
this term could be neglected within the long time regime. However,
as we will show below, there is no need to introduce approximations
because it will cancel out naturally when we introduce the initial
conditions of the problem. The effective action for a single
particle coupled to the TLSs reservoir then reads \begin{eqnarray}
S_{\textrm{eff}}[q,\xi ] & = & \int _{0}^{t}dt'\left[M\dot{q}(t')\dot{\xi }(t')-q(0){\tilde{\Gamma }}(t')\xi (t')\right.\nonumber \\
 &  & \left.-\int _{0}^{t'}dt''{\tilde{\Gamma }}(t'-t'')\dot{q}(t'')\xi (t')\right].\label{eq:actdef}
\end{eqnarray}

Before explicitly solving the equation of motion corresponding to
the action (\ref{eq:actdef}), it is convenient to specify the spectral
density of the thermal bath in terms of macroscopic parameters. A
reasonable assumption\cite{twlyo} for it is \begin{equation}
J(\omega ,T)=\frac{\eta }{\pi }\, \left(\frac{\omega }{\omega _{c}}\right)^{s}\tanh \left(\frac{\hbar \omega }{2k_{B}T}\right)\; \Theta (\Omega -\omega ),\label{eq:spectral}\end{equation}
 where $\Omega $ is a cutoff frequency, $\eta $ is a constant defining
the coupling strength of the particle to the TLSs, $s$ is a number
(real and positive) which determines the long time properties of the
thermal bath, and $\omega _{c}$ is some characteristic frequency
introduced in order to make the unit of $\eta $ independent of $s$.
Notice that the temperature dependence in Eq.\ (\ref{eq:spectral})
is crucial for a fermionic heat bath because it ensures that the bath
degrees of freedom are excited as the temperature increases.\cite{comentario}
In Ref.\ {[}\onlinecite{Hjing}{]} this point was not acknowledged,
leading to the wrong conclusion that the decaying term in the particle
equation of motion is temperature independent.

Using Eqs.\ (\ref{eq:actdef}) and (\ref{eq:spectral}), the classical
equations of motion for $q$ and $\xi $ can be written as \begin{equation}
\ddot{q}+\frac{2\gamma }{\pi }\int _{0}^{t}dt'\Gamma (t-t')\, \dot{q}(t')+\frac{2\gamma }{\pi }q(0)\Gamma (t)=\frac{eE(t)}{M},\label{eq:equaq}\end{equation}
\begin{equation}
\ddot{\xi }-\frac{2\gamma }{\pi }\int _{0}^{t}dt'\Gamma (t-t')\, \dot{\xi }(t')+\frac{2\gamma }{\pi }\xi (0)\Gamma (-t)=0,\label{eq:equaxi}\end{equation}
 where the damping constant is defined as $\gamma =\eta /2M$ and
the kernel $\Gamma $ is given by \begin{equation}
\Gamma (t)=\int _{0}^{\Omega }d\omega \frac{\omega ^{s-1}}{\omega _{c}^{s}}\, \tanh \left(\frac{{\hbar \omega }}{2k_{B}T}\right)\cos (\omega t).\label{eq:gamma}\end{equation}
 Therefore, after tracing out the TLSs reservoir we obtained an equation
of motion for the particle center of mass (\ref{eq:equaq}) in which
the thermal bath has the same effect as that of a \emph{viscous fluid}.

For any value of $s$ the solution of Eqs.\ (\ref{eq:equaq}) and
(\ref{eq:equaxi}) can be written in terms of the kernel Laplace transform
as\begin{equation}
\overline{q}(z)=\frac{{zq(0)+\dot{q}(0)}}{z²+2\gamma z\Gamma (z)/\pi },\label{eq:qzsol}\end{equation}
\begin{equation}
\overline{\xi }(z)=\frac{{z\xi (0)+\dot{{\xi }}(0)}}{z²-2\gamma z\Gamma (z)/\pi },\label{eq:xizsol}\end{equation}
 where\begin{eqnarray}
\Gamma (z) & = & \frac{\Omega ^{s+1}}{s+1}\left[\frac{{}_{2}F_{1}(1,\frac{1+s}{2},\frac{3+s}{2},-\frac{\Omega ²}{z²})}{z²}\tan \left(\frac{\hbar z}{2kT}\right)\right.\nonumber \\
 &  & \left.-\frac{4k_{B}Tz}{\hbar }\sum _{n=1}^{\infty }\frac{\, _{2}F_{1}(1,\frac{1+s}{2},\frac{3+s}{2},-\frac{\Omega ²}{\lambda _{n}²})}{\lambda _{n}²(\lambda _{n}^{2}-z²)}\right].\label{eq:hyper1}
\end{eqnarray}
 In the expression above $_{2}F_{1}$ denotes the hyper-geometric
function and $\lambda _{n}=(2n-1)\pi k_{B}T/\hbar $, with $n\in \mathbb{N}$.
It should be notice that the fluctuating force, given by the last
term on the LHS in Eq.\ (\ref{eq:equaq}), does not appear in Eq.\ (\ref{eq:qzsol}).
This term was exactly canceled by the initial condition included in
the Laplace transform of the damping term and therefore there is no
need to drop it out by assuming the long time approximation. In order
to illustrate how the dissipative properties of the TLSs thermal reservoir
affects the single particle dynamics in the \emph{sub-ohmic} regime
($s<1$), lets investigate the simple case in which $q(0)=0$ and
$\dot{q}(0)=v_{0}$.

\section{The {}``dynamical'' localization}

We start by discussing the $s=0$ case, in which
we can proceed analytically a bit further. In this case the hyper-geometric
function reads $_{2}F_{1}(1,1/2,3/2,-x²)=x^{-1}\arctan x$ and the
Laplace transform of the damping function given by Eq.\ (\ref{eq:hyper1})
acquires the form \begin{equation}
\Gamma =\sum _{n=1}^{\infty }\frac{4k_{B}Tz}{\hbar (\lambda _{n}^{2}-z^{2})}
\left[\frac{\tan ^{-1}\left(\frac{\Omega }{z}\right)}{z}-\frac{\tan ^{-1}
\left(\frac{\Omega }{\lambda _{n}}\right)}{\lambda _{n}}\right].
\label{eq:ltrax}
\end{equation}

In the particular case of $\Omega \rightarrow \infty $, the high
temperature limit of Eq.\ (\ref{eq:ltrax}) is $\Gamma (z)=\hbar /2k_{B}T$
and therefore the effective dynamics of the particle of interest simply
becomes \begin{equation}
q(t)=\frac{v_{0}}{\hbar \gamma /2k_{B}T}(1-e^{-\frac{\hbar \gamma }{2k_{B}T}t}).\label{htlimit}\end{equation}
 This result correctly reproduces the oscillator-bath model with an
ohmic temperature dependent damping constant. Indeed, if we assume
$\hbar \omega \ll k_{B}T$ and $\Omega \rightarrow \infty $ in Eq.\ 
(\ref{eq:gamma}) the damping function becomes $\Gamma (t-t')=(\hbar \pi /2k_{B}T)\delta (t-t')$,
which is an instantaneous function, indicating that the dissipative
process is completely memoryless and the condition of zero frictional force is achived only 
when the particle velocity is zero.
Physically, this limit corresponds
to a weak particle-reservoir interaction because most of the TLSs
are occupied (on average), causing no damping on the particle. Therefore,
we recover the known result demonstrated in Ref.\ {}{[}\onlinecite{FeyVer}{]},
namely that when the coupling between the particle of interest and
a nonlinear bath is weak enough, the latter behaves as a collection
of harmonic oscillators.

\begin{figure}
\begin{center} \includegraphics[clip, scale=0.30]{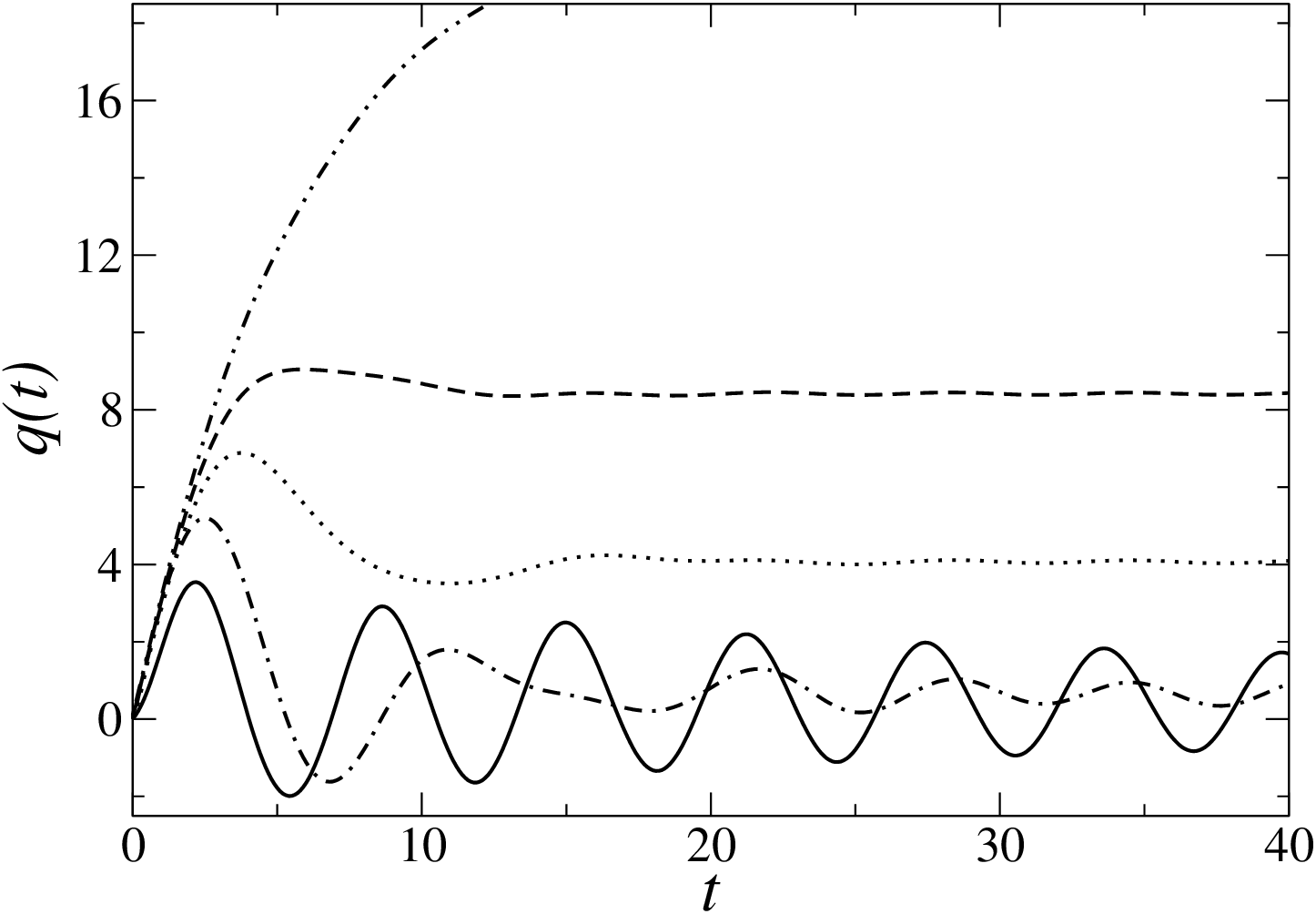}\end{center}

\caption{\label{tdepS0}Time dependence of the particle center of mass for
$s=0$ and different values of temperature. The continuous line corresponds
to $T=0.0001$, while the dash-dotted, dotted, dashed and double-dotted
lines are for $T=0.01$, $T=0.1$, $T=0.2$ and $T=0.5$, respectively.
In all cases $\Omega =1$, $\gamma =0.3$ and the initial velocity
was taken equal to $1$.}
\end{figure}

However, the features discussed above are not valid for all values
of temperature and frequency cutoff. In fact, we realize that at $T=0$,
even assuming $\Omega \rightarrow \infty $, it is impossible to obtain
a damping function without memory. In this case the problem becomes
non-Markovian and although the particle becomes localized after some
time, neither its position nor its velocity ever reach the equilibrium.
This behavior is illustrated in Fig.\ref{tdepS0} for finite $\Omega $
and different values of temperature. It is worth to notice that in
this situation the particle dynamics completely differs from its behavior
in the oscillator bath model. In the later, at zero velocity, the
frictional force over the particle is also zero and classically, the
particle remains in that state forever. In our non-Markovian situation,
the frictional force acting on the particle depends on the previous
velocities with different weights - given by the kernel (\ref{eq:gamma})
- and the situation of zero frictional force over the particle, at
a given instant, does not correspond to zero velocity. 

It is not difficult
to see that at low temperature, the kernel oscillates
in time, keeping its sign constant and therefore the only possibility
of getting zero frictional force acting on the particle at
a given instant occurs when the particle changes the momentum direction
within the time interval. From the physical point of view, this behavior
resembles that of a particle confined by a potential. Indeed, the
particle oscillation around this effective potential, which clearly
appears for $T=0.0001$ in Fig.\ \ref{tdepS0}, can be promptly obtained
within the long time regime. The term $\cos [\omega (t-t')]$ in Eq.\ (\ref{eq:gamma})
then oscillates rapidly, yielding no contribution to the damping process
for long times, except when $\omega \ll (t-t')^{-1}$. In such a situation,
$\Gamma (t)$ becomes finite and time independent, turning the damping
term into a harmonic localizing potential. As the temperature increases,
less reservoir states are able to play their dissipative role and
the particle takes longer to get localized far from the origin.

We can therefore conclude that the particle dynamics in this situation
($s=0$) is completely different from the one obtained when the thermal
bath is represented by the usual oscillator model. Here, memory effects
in the damping process lead to a ``dynamical'' localization of
the particle at a certain distance from the initial position, which
is proportional to the temperature. The strength of the localization
potential is determined by the ratio of two quantities, namely the
thermal and the cutoff energies. This quantity measures the number
of reservoir states which effectively couple to the particle and obviously
also depends on the total reservoir states determined by the spectral
function.

\begin{figure}
\begin{center}\includegraphics[  clip,
  scale=0.30]{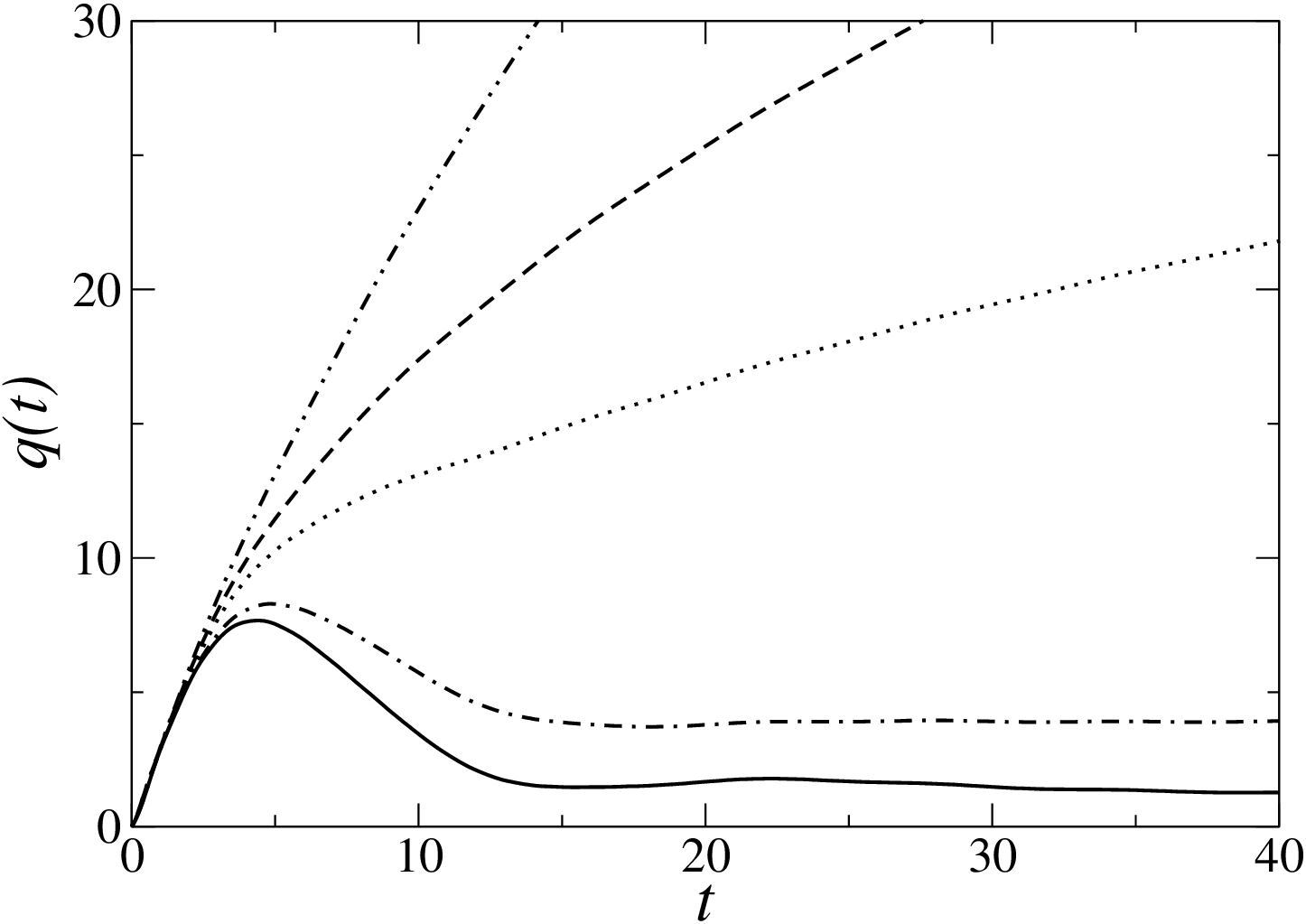}\end{center}

\caption{\label{tdepS05}Time dependence of the particle center of mass for
$s=0.5$ and different values of temperature. The continuous line
corresponds to $T=0.0001$, while the dash-dotted, dotted, dashed
and double-dotted lines are for $T=0.01$, $T=0.1$, $T=0.2$ and
$T=0.5$, respectively. In all cases $\Omega =1$, $\gamma =0.3$ and
the initial velocity was taken equal to $1$.}
\end{figure}

\begin{figure}
\begin{center}\includegraphics[  clip,  scale=0.30]{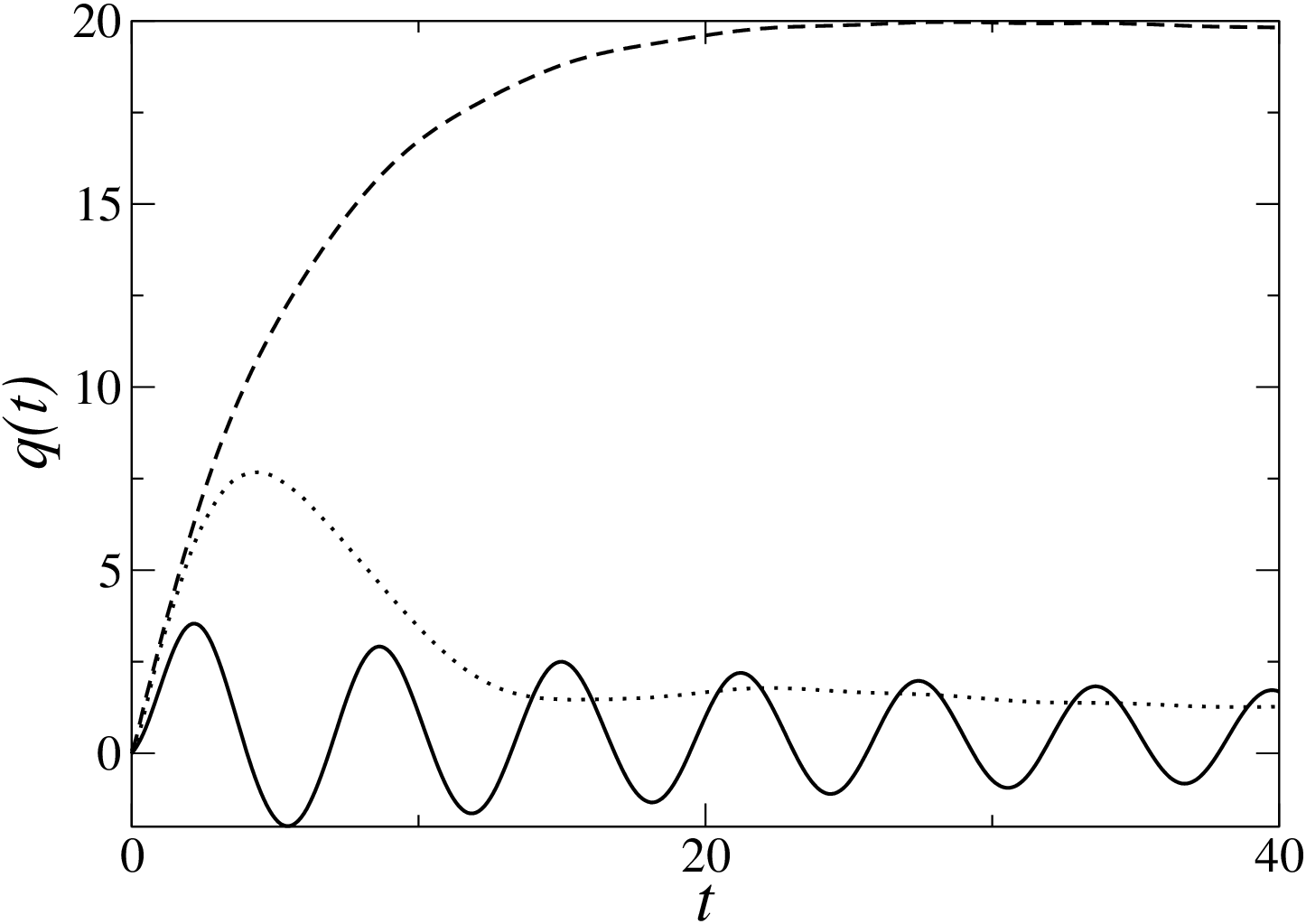}\end{center}

\caption{\label{sdiff}Time dependence of the particle center of mass for
$T=0.001$ and different values of $s$. The continuous line corresponds
to $s=0$, while the dotted and dashed lines are for $s=0.5$ and
$s=1$, respectively. In all cases $\Omega =1$, $\gamma =0.3$ and
the initial velocity was taken equal to $1$.}
\end{figure}

In order to get more insight on the particle dynamics in the sub-ohmic
regime we have plotted in Fig. \ref{tdepS05} the position as a function
of time in the specific case of $s=0.5$. At very low temperatures,
the main difference from the $s=0$ case is that the localizing potential
strength in the long time regime becomes weaker and the dynamical
localization effect is difficult to observe, see the $T=0.0001$ and
$T=0.01$ cases, for instance. This is a consequence of having decreased
the weight of the low energy reservoir states in the spectral function.
At the same time, as the temperature increases, the number of low
energy states effectively contributing to the dissipative process
becomes reduced because several low-energy states are occupied (on
average) and the particle moves nearly free, see Fig.\ \ref{tdepS05}
for $T=0.2$ and $T=0.5$. This last behavior differs from the $s=0$
situation in which, even at high temperatures, there are enough low
energy states to localize the particle. In general, the particle dynamics
in the sub-ohmic regime will be described by a function which exhibits
a behavior in between that of the $s=0$ and $s=1$ limiting cases.
This behavior is illustrated for low temperatures in Fig.\ \ref{sdiff}.
The central point in the ohmic case $(s = 1)$ is that even at low temperatures,
there are not enough low energy states that render the particle confinement
appreciable in the long time regime.

\section{Conclusions}

We have studied the real time dynamics of a particle coupled
to a TLSs thermal reservoir and found that within the sub-ohmic
regime the particle becomes localized in the long time limit, oscillating
in the real space as a consequence of an effective potential generated
by its interaction with the thermal bath. The oscillatory behavior
renders the localization {}``dynamical'' and therefore neither the
particle position nor its velocity ever reach the equilibrium. 
This behavior is associated with the non-Markovian character of the
dissipative process, which in our simple model is provided by 
inelastic scattering of the particle of interest by the TLSs. 
We hope that our findings
can be of some help in the understanding of transport properties of
systems in which the dissipative medium seems to be sub-ohmic.\cite{experi,sachdev,dudu}
We also speculate about the extension of the model discussed here
to a situation in which the fermionic bath excitation has a finite
gap $\Delta (T)$. In such a case, we expect that the effect of {}``dynamical''
localization will start at temperatures above $\Delta (T)/k_{B}$,
which is probably more appropriate to describe realistic situations.

\end{document}